\documentclass[graybox]{svmult}

\usepackage{helvet}         % selects Helvetica as sans-serif font
\usepackage{courier}        % selects Courier as typewriter font
\usepackage{type1cm}        % activate if the above 3 fonts are
\usepackage{makeidx}         % allows index generation
\usepackage{graphicx}        % standard LaTeX graphics tool
\usepackage{multicol}        % used for the two-column index
\usepackage[bottom]{footmisc}% places footnotes at page bottom

\usepackage{bm}
\usepackage{dcolumn}
\usepackage{color}
\usepackage{braket} 
\usepackage{amsmath}
\usepackage{amsbsy}

\makeindex             % used for the subject index
\begin{document}

\title*{Spin interference effects in Rashba quantum rings}
\author{Carmine Ortix}
\institute{Carmine Ortix \at Institute for Theoretical Physics, Center for Extreme Matter and Emergent Phenomena, Utrecht University, Princetonplein 5, 3584 CC Utrecht, The Netherlands; \\ Dipartimento di Fisica ``E. R. Caianiello", Universit\'a di Salerno, IT-84084 Fisciano, Italy \\ \email{c.ortix@uu.nl}}
%\and Name of Second Author \at Name, Address of Institute \email{name@email.address}}
\maketitle

\abstract{Quantum interference effects in rings provide suitable means to control spins at the mesoscopic scale. In this chapter we present the theory underlying spin-induced modulations of unpolarized currents in quantum rings subject to the Rashba spin-orbit interaction. We discuss explicitly the connection between the conductance modulations and the geometric phase acquired by the spin during transport, as well as pathways to directly control them. }

\section{Quantum rings with Rashba spin-orbit interaction: effective one-dimensional Hamiltonian}
\label{sec:hamiltonian}
The effect of the Rashba spin-orbit interaction~\cite{ras60} on electrons moving in mesoscopic rings has been studied in several contexts, including magnetoconductance oscillations~\cite{aro93,yi97} and persistence currents~\cite{zhu95,qia97}. Essentially all these theoretical studies have employed one-dimensional (1D) model Hamiltonians. Different Hamiltonians have been used by different authors in the past, and consequently some ambiguity with regard to the correct form of the 1D Hamiltonian exists in the literature. Aronov and Lyanda-Geller~\cite{aro93}, for instance, studied the effect of the Rashba spin-orbit interaction on the Aharonov-Bohm conductance oscillations using a non-Hermitean operator. 
The procedure for obtaining the correct one-dimensional Hamiltonian in quantum rings in the presence of Rashba spin-orbit interaction has been first provided by Meijer, Morpurgo, and Klapwijk~\cite{mei02}, who started out from the full two-dimensional (2D) Hamiltonian of a two-dimensional electron gas (2DEG) subject to a strong confining potential with circular symmetry forcing the electrons to be localized on the quantum ring in the radial direction. This procedure, which is the most rigorous, and physically sound one, corresponds precisely to the so-called ``thin-wall" quantization procedure originally introduced by Jensen, Koppe~\cite{jen71}, and Da Costa (JKC)~\cite{dac82} 
to describe the quantum mechanics of non-relativistic particles constrained to generic ``curved" $n$-dimensional manifolds but embedded in a $n+1$ Euclidean space. 

In the absence of spin-orbit coupling, the JKC approach predicts the existence of a curvature-induced quantum geometric potential (QGP), which causes intriguing phenomena at the nanoscale \cite{can00,aok01,enc03,fuj05,kos05,gra05,cha04,mar05,ved00,cuo09,ort10}.
In periodically minimal surfaces, for instance, the QGP leads to a topological band structure \cite{aok01}. Similarly, in spirally rolled-up nanotubes the QGP has been shown to lead to winding-generated bound states \cite{ort10}. These curvature effects have been predicted to become even more pervasive in strain-driven nanostructures where the nanoscale variation of strain induced by curvature leads to a strain-induced geometric potential that is of the same functional form as the QGP, but gigantically boosting it \cite{ort11b}. 

The JKC thin-wall approach has been recently shown to be well founded also in presence of externally applied electric and magnetic fields \cite{fer08,ort11} and subsequently employed to predict novel curvature-induced phenomena, such as the strongly anisotropic ballistic magnetoresistance of spirally rolled-up semiconducting nanotubes without magnetism and spin-orbit interaction \cite{cha14}. Finally, the experimental realization of an optical analog of the curvature-induced QGP has provided  empirical evidence for the validity of the JKC squeezing procedure \cite{sza10}. 
As we will show below, the JKC procedure can be also applied without restrictions in the presence of spin-orbit coupling, thereby allowing to derive the correct Hermitean Hamiltonian of quantum rings with an arbitrary geometric shape. 

To start with, we recall that in the usual effective-mass approximation, the movement of the charge carriers in presence of spin-orbit interaction can be described with an effective Schr\"odinger-Pauli equation acting on  a two-dimensional spinor $\psi$:
\begin{equation}
\left(\dfrac{{\bf p}^2}{2 m^{\star}} + \, {\boldsymbol \alpha} \cdot {\boldsymbol \sigma} \times {\bf p} \right) \psi = E \, \psi, 
\label{eq:ham0}
\end{equation}
where  ${\bf p} = -i \hbar \nabla$ is the canonical momentum operator and the $\boldsymbol{\sigma}$'s are the usual Pauli matrices generating the Clifford algebra of ${\mathcal R}^3$ , which obey the anticommutation relations $\left\{ \sigma_i, \sigma_j \right\} = 2 \, \eta_{i j}$ with  $\eta_{i j}$  the standard  spatial metric given by the identity matrix. In addition, we introduced  the vector ${\boldsymbol \alpha}$ with magnitude corresponding to the  spin-orbit interaction constant, and direction determined by the effective electric field  from which the spin-orbit coupling originates.  
Finally $m^{\star}$ is the material dependent effective mass of the carriers. 
 In the remainder, we will use Latin indices for spatial tensor components of the flat Euclidean three-dimensional space whereas Greek indices will be used for the corresponding tensor components in curved space. 
Adopting Einstein summation convention, Eq.~\ref{eq:ham0} can be generalized to a curved three-dimensional  space as follows 
\begin{eqnarray}
E \psi&=& \left[ - \frac{\hbar^2}{ 2 m^{\star}} \left(G^{\mu \nu} \partial_\mu \partial_\nu - G^{\mu \nu} \,  \Gamma_{\mu \nu}^{\lambda} \partial_\lambda \right) \right.\nonumber \\ & & \left. -i \, \hbar \, \, {\mathcal E}^{\mu \nu \lambda}  \, \alpha_{\mu} \varsigma_{\nu} \partial_{\lambda} \right] \psi,
\label{eq:hamcurved}
\end{eqnarray}
where $G^{\mu \nu}$ is the inverse of the metric tensor $G_{\mu \nu}$, ${\mathcal E}^{\mu \nu \lambda}$ is the contravariant Levi-Civita tensor -- it can be written in terms of the usual Levi-Civita symbol as ${\mathcal E}^{\mu \nu \lambda} = \epsilon^{\mu \nu \lambda} / \sqrt{ || G || }$ -- and we introduced the affine connection
$$\Gamma_{\mu \nu}^{\lambda} = \dfrac{1}{2} G^{\lambda \xi} \left[\partial_\nu G_{\xi \mu} + \partial_{\mu} G_{\xi \nu} - \partial_{\xi} G_{\mu \nu} \right]. $$ 
Finally, the $\varsigma$'s are the generators of the Clifford algebra in curved space $\left\{\varsigma_\mu , \varsigma_{\nu} \right\} = 2 \, G_{\mu \nu}$.   

To proceed further, we need to define a coordinate system. We therefore start out by defining a planar curve ${\mathcal C}$ of parametric equations ${\bf r}={\bf r}(s)$ with $s$ indicating the corresponding arclength. The portion of the three-dimensional space in the immediate neighborhood of ${\mathcal C}$ can be then parametrized as ${\bf R}(s ,q_2,q_3) = {\bf r}(s) +  \hat{N}(s) \, q_2 + \hat{B} \, q_3,$
where $\hat{N}$ is the unit vector normal to ${\cal C}$, but residing in the curve plane, while $\hat{B}$ is the binormal vector perpendicular to the quantum ring plane. 
The structure of the corresponding three-dimensional spatial metric tensor can be determined using that the two orthonormal vectors ${\hat T}(s) = \partial_s {\bf r}(s)$ and ${\hat N}(s)$ obey the Frenet-Serret type equations of motion as they propagate along $s$
\begin{equation}
\left(
\begin{array}{c} \partial_s {\hat T}(s) \\ \partial_s {\hat N}(s) 
\end{array}
\right) = \left( \begin{array}{cc} 0 & \kappa(s)  \\ 
-\kappa(s) & 0 
\end{array}
\right)
\left( \begin{array}{c}  {\hat T}(s) \\  {\hat N}(s) \end{array}  \right),
\end{equation}
where $\kappa(s)$ denotes the local curvature of the quantum ring. With this, the metric tensor corresponding to the three-dimensional portion of space explicitly assumes the diagonal form
\begin{equation*}
G=\left( \begin{array}{ccc} \left[1 - \kappa(s) q_2 \right]^2  & 0 & 0 \\
0 & 1 & 0 \\ 
0 & 0 & 1 \end{array} \right),
\end{equation*}
whose determinant $||G||=\left[1 - \kappa(s) q_2 \right]^2$. The generators of the Clifford algebra for the metric tensor written above can be derived introducing the Cartan's dreibein formalism \cite{car04}.  At each point, we define  a set of one forms  with components $e_{\mu}^i$ and  a dual set of vector fields $e_i^{\mu}$ obeying the duality relations $e_\mu^i e^{\nu}_i = \delta^{\mu}_{\nu}$ and $e_{\mu}^i e_{j}^{\nu}= \delta_{i}^j$, and corresponding to the "square root" of the metric tensor $G_{\mu \nu} = e_{\mu}^{i} \delta_{i j } e_{\nu}^j$. The generators of the Clifford algebra can be then expressed as $\varsigma_\mu = e_\mu^i \sigma_i$. For the metric tensor written above, the dreibein field can be chosen as 
$e_{s}^{i} = \hat{T}^i(s) \left(1 - \kappa(s) q_2 \right)$,  $e_{q_2}^i= \hat{N}^i(s)$ and $e_{q_3}^i =  \hat{B}^i(s)$.  This immediately allows to identify the $\varsigma$'s  as $\varsigma_s = \sigma_T  \left(1 - \kappa(s) q_2 \right)$, $\varsigma_{q_2} = \sigma_N$, and $\varsigma_{q_3}= \sigma_B $ written in terms of a local set of three Pauli matrices comoving  with the Frenet-Serret frame $\sigma_{T,N,B}= \boldsymbol{\sigma} \cdot (\hat{T}, \hat{N},\hat{B}$).  

In the same spirit of JKC \cite{jen71,dac82}, we now apply a thin-wall quantization procedure and take explicitly into account the effect of two strong confining potentials in the normal and binormal directions $V_{\lambda_N}(q_2)$, $V_{\lambda_B}(q_3)$ respectively, with $\lambda_{N,B}$ the two independent squeezing parameters. Furthermore, we introduce a rescaled spinorial wavefunction $\chi$ such that the line probability can be defined as $\int \chi^{\dagger} \chi \, d q_2 \, d q_3$. Conservation of the norm requires 
$${\cal N} = \int \sqrt{||G||} \, d s \, d q_2 \, d q_3 \, \psi^{\dagger} \psi =  \int d s \, d q_2\,  d q_3 \, \chi^{\dagger} \chi,$$
from which the rescaled spinor $\chi \equiv \psi \times ||G||^{1/4}$.

In the $\lambda_{N,B} \rightarrow \infty$ limit, the spinorial wavefunction will be localized in a narrow range close to $q_{2,3}=0$. This allows us to expand all terms appearing in Eq.~\ref{eq:hamcurved}  in powers of $q_{2,3}$. At the zeroth order we then obtain the following Schr\"odinger-Pauli equation: 
\begin{eqnarray}
E \, \chi&=& \left[ - \dfrac{\hbar^2}{ 2 m^{\star}} \left(\eta^{\mu \nu} \partial_\mu \partial_\nu + \dfrac{\kappa(s)^2}{4}  \right) -i  \hbar \, \, {\epsilon}^{\mu \nu \lambda}  \, \alpha_{\mu} \sigma_{\nu} \partial_{\lambda} \right.  \nonumber \\  & & \left.  -i \hbar \, \, {\epsilon}^{\mu \nu q_2}  \, \alpha_{\mu} \sigma_{\nu} \dfrac{\kappa(s)}{2} + V_{\lambda_N}(q_2) + V_{\lambda_B}(q_3)  \right] \chi
\label{eq:hamcurved2}
\end{eqnarray}
In the equation above, we have used that in the $q_{2,3} \rightarrow 0$ limit  the only non-vanishing affine connection component $\Gamma_{s \, s}^{q_2} = \kappa(s)$, and employed the limiting relations for the derivatives of the original spinor in terms of the rescaled one
\begin{equation*} 
\left\{ \begin{array}{l} \partial_{q_{2}} \psi = \partial_{q_2} \chi +\dfrac{\kappa(s)}{2} \chi  \\  \\
  \partial^2_{q_{2}} \psi = \partial^2_{q_2} \chi + \kappa(s) \partial_{q_2} \chi + \dfrac{3}{4} \kappa(s)^2 \chi.
\end{array} \right.
\end{equation*}
The presence of the relativistic spin-orbit interaction in Eq.~\ref{eq:hamcurved2} prevents  the separability of the quantum dynamics along the tangential direction of the planar curve from the normal  quantum motion. However, the strong size quantization along the latter direction still allows us to employ an adiabatic approximation \cite{ort11b}, encoded in the ansatz for the spinorial wavefunction $\chi(s,q_2,q_3)=\chi_T (s) \times \chi_N (q_2) \times \chi_B (q_3)$ where the normal and binormal wavefunctions solve the Schr\"odinger equation
$$-\frac{\hbar^2}{2 m^{\star}} \partial^2_{q_{2}, q_{3}} \, \, \chi_{N, B} + V_{\lambda_{N,B}} (q_{2,3}) \, \chi_{N,B} = E_{N,B} \, \chi_{N,B}.$$
We can assume the two confining potential to take either the form of an harmonic trap $\propto q_{2,3}^2$ or an infinite potential well centered at $q_{2,3} \equiv 0$. Taken perturbatively, the first derivatives terms $\partial_{q_{2,3}}$ of Eq.~\ref{eq:hamcurved2} vanish and thus the effective one-dimensional Schr\"odinger-Pauli equation for the tangential wavefunction reads
\begin{eqnarray}
E \, \chi_T &=&  \left[ - \dfrac{\hbar^2}{ 2 m^{\star}} \left(\partial^2_{s} + \dfrac{\kappa(s)^2}{4}  \right) -i \hbar \alpha_N \sigma_B \partial_s \right. \label{eq:hamcurved1D1} \\ & & \left. + i \hbar \alpha_B \left( \sigma_N \partial_s - \sigma_T \dfrac{\kappa(s)}{2} \right) \right] \chi_T,
\nonumber
\end{eqnarray}
where we explicitly considered a spin-orbit coupling originating either from an electric field orthogonal to the ring plane ($\alpha_{N}$) or from an electric field pointing in the normal direction to the ring ($\alpha_{B}$). 
Eq.~\ref{eq:hamcurved1D1} 
represents the correct effective one-dimensional Schr\"odinger-Pauli equation for a single electron in presence of Rashba spin-orbit interaction, and generalizes the result obtained for a circular quantum ring \cite{mei02,fru04,gen13}. 
The corresponding Schr\"odinger-Pauli operator is indeed Hermitian as can be shown by calculating its matrix elements in any complete basis, or simply noticing that 
it can be written, using anticommutators, in the compact form 
\begin{eqnarray*}
E \chi_T &=& \left[ \dfrac{\hat{p}_s^2}{2 m^{\star}} - \dfrac{\hbar^2 \kappa(s)^2 }{8 m^{\star}} +\dfrac{\alpha_N}{2} \left\{\hat{p}_s , \sigma_B \right\} \right. \\ & & \left.  -\dfrac{\alpha_B}{2} \left\{{\hat p}_s , \sigma_N \right\} \right] \chi_T, 
\end{eqnarray*}
where the tangential momentum operator ${\hat p}_s=-i \hbar \partial_s$.

\section{Conductance modulations in Rashba circular quantum rings} 
\label{sec:circularring}
\begin{figure}[t]
\begin{center}
\includegraphics[width=.75\columnwidth]{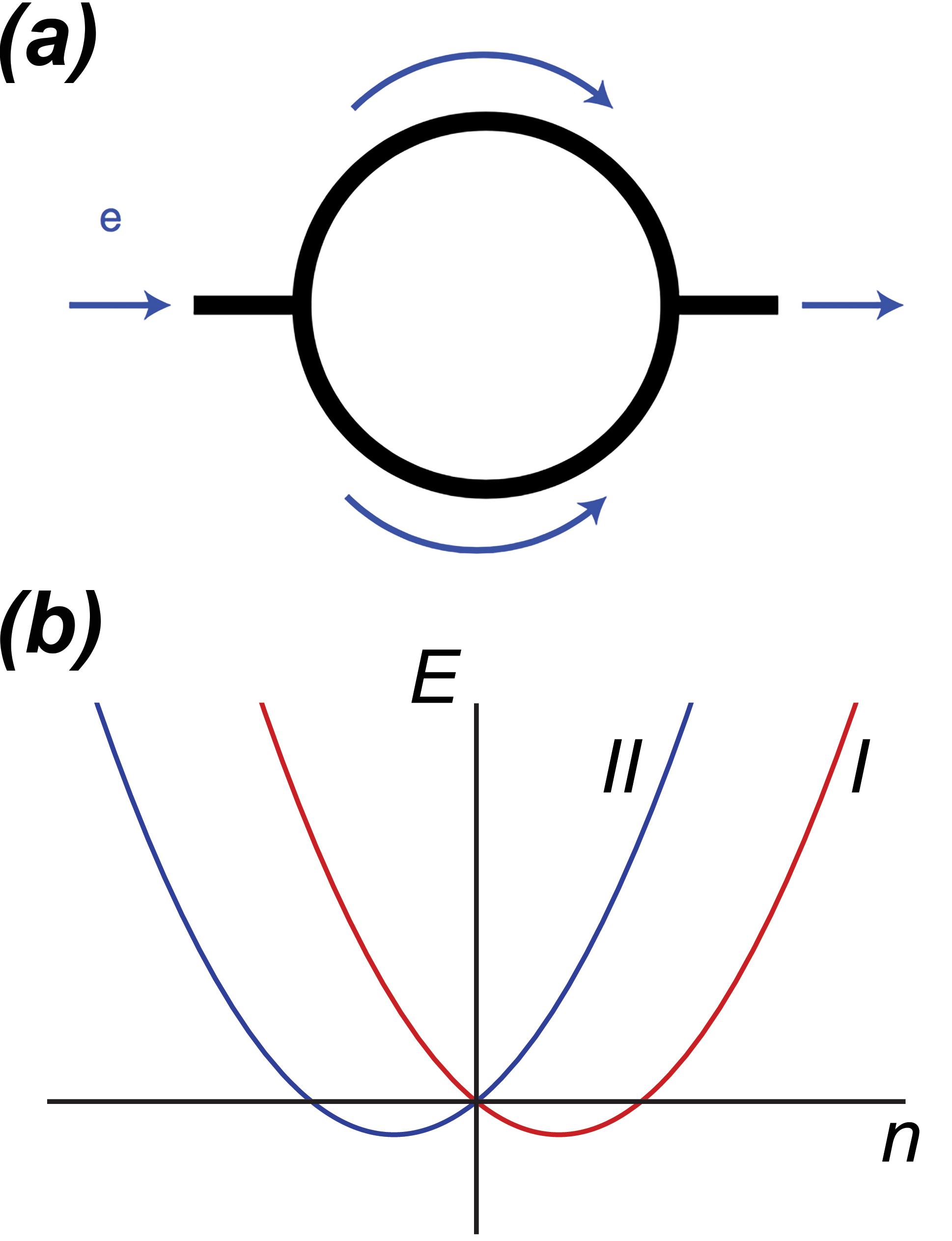}
\end{center}
\caption{(a) Sketch of the spin interferometer devices based on a mesoscopic ring with Rashba spin-orbit interaction (Adapted from Ref.~\cite{nag13}). (b) Energy levels in a quantum ring with Rashba spin-orbit interaction plotted as a function of the mode quantum number $n$. The two time-reversal channels $I$,$II$ are indicated.}
\label{fig:fig1}
\end{figure}
In this section, we discuss the quantum transport properties of a mesoscopic ballistic device in which a circular quantum ring with Rashba spin-orbit couplings is  symmetrically coupled to two contact leads [c.f. Fig.~\ref{fig:fig1}(a)]. 
The transport properties can be analyzed straightforwardly in the linear response regime, in which the system is subject to a constant, low-bias voltage. 
According to the Landauer formula, the  zero-temperature conductance reads~\cite{fru04}
\begin{equation}
G=\dfrac{e^2}{h} \sum_{m, m^{\prime}=1}^{N} \sum_{\sigma \sigma^{\prime}} T_{m^{\prime}, m}^{\sigma^{\prime}, \sigma}
\end{equation}
where $T_{m^{\prime}, m}^{\sigma^{\prime}, \sigma}$ denotes the quantum probability of transmission between incoming $(m, \sigma)$ and outgoing $(m^{\prime}, \sigma^{\prime})$ states on the semi-infinite ballistic leads, with $m,m^{\prime}$ and $\sigma, \sigma^{\prime}$  the mode and spin quantum numbers, respectively. 
The total number of modes $M=1$ in an effective one-dimensional description. 
Assuming perfect couplings between the leads and the ring, and thus neglecting backscattering effects, the quantum transmission probability are entirely determined by the eigenstates of the Hamiltonian Eq.~\ref{eq:hamcurved1D1} for the Rashba spin-orbit coupled quantum ring.

We first analyze a circular quantum ring with a Rashba spin-orbit interaction due to a radial electric field~\cite{mat92,cho97}. Adopting polar coordinates, the effective one-dimensional Hamiltonian Eq.~\ref{eq:hamcurved1D1} then takes the following form: 
\begin{equation}
{\cal H}=-\dfrac{\hbar^2}{2 m^{\star} R^2} \partial^2_{\phi} + i \dfrac{\alpha \hbar}{R} \sigma_z \partial_\phi.
\end{equation}
The corresponding one-dimensional spinorial eigenstates can be simply found as 
\begin{equation*}
\Psi^{\uparrow}_{n}(\phi)=\mathrm{e}^{i n \phi} \left( \begin{array}{c} 1 \\ 0 \end{array} \right), 
\end{equation*}
\begin{equation*}
\Psi^{\downarrow}_{n}(\phi)=\mathrm{e}^{i n \phi} \left( \begin{array}{c} 0 \\ 1 \end{array} \right), 
\end{equation*}
with the associated eigenenergies reading 
$$E_{\uparrow, \downarrow}(n) = \dfrac{\hbar^2}{2 m^{\star} R^2} n^2 \mp \dfrac{ \hbar \alpha}{R} n. $$
The energy splitting due to the Rashba spin-orbit interaction implies that incoming spins $\ket{\sigma}$ entering the ring at $\phi=0$ with a Fermi energy $E_F$ can propagate coherently along four different channels obtained by solving $E_{\uparrow,\downarrow}(n) \equiv E_F$. Specifically, two opposite spin states $\ket{n_1 ; \uparrow}$, $\ket{n_2 ; \downarrow}$ propagate along the upper branch of the ring, whereas their time-reversal partners $\ket{-n_1 ; \downarrow}$, $\ket{-n_2 ; \uparrow}$ propagate along the lower branch of the ring. The interference between the channels at $\phi=\pi$ then implies that injected spins leave the ring in a mixed spin state: 
$$\ket{\sigma_{out}} = \sum_{s= \uparrow,\downarrow} \sum_{i=1,2} \braket{n_i ; s | \sigma} \times \mathrm{e}^{i n_i \pi} \ket{n_{i}; s}. $$
Choosing a complete basis of incoming and outgoing spin states, the spin-resolved transmission probabilities are obtained as $T^{\sigma^{\prime} \sigma} = |\braket{\sigma^{\prime} | \sigma_{out}}|^2$. By further summing over the spin indices $\sigma^{\prime}$ and $\sigma$, we thereby obtain the total conductance
\begin{equation}
G= \dfrac{e^2}{h} \left[1 + \cos{(n_1 - n_2) \pi}\right]
\label{eq:conductance1}
\end{equation}
The relation between the two wave numbers $n_{1},n_{2}$ can be simply found to be $n_1- n_2 \equiv Q_{R} \equiv 2 m^{\star} R \,  \alpha / \hbar$. With this, it follows that the conductance exhibits uniform oscillations as a function of the spin-orbit interaction strength, which is the signature of the Aharonov-Casher effect~\cite{aha84} for spins traveling in an external electric field . 

The radial electric field considered above, however,  does not correspond to the normal situation in which the electric field is orthogonal to the plane in which the quantum ring resides. 
When considering this, the one-dimensional Hamiltonian Eq.~\ref{eq:hamcurved1D1} for a quantum ring with circular symmetry explicitly reads: 
\begin{equation}
{\cal H}=-\dfrac{\hbar^2}{2 m^{\star} R^2} \partial^2_{\phi} + i \dfrac{\alpha \hbar}{R} \left[\sigma_N \partial_\phi + \dfrac{\sigma_T}{2} \right],
\label{eq:hamiltonianringreal}
\end{equation}
where we introduced the two local Pauli matrices 
\begin{equation}
\left\{ \begin{array}{ccc} \sigma_N &=& \cos{\phi}\, \sigma_x + \sin{\phi}\, \sigma_y \\ 
\sigma_{T}&=& -\sin{\phi}\, \sigma_x + \cos{\phi}\, \sigma_{y} 
\end{array}
\right. .
\end{equation}
The spinorial eigenstates of the Hamiltonian above can be found using a trial spinorial wavefunction of the form $\Psi= \mathrm{e}^{i n \phi} \times \left[\chi_1 \mathrm{e}^{-i \phi/2}, \chi_2 \mathrm{e}^{i \phi / 2} \right]^T$, where the amplitudes $\chi_{1,2}$ are determined by the effective Hamiltonian 
\begin{equation}
\widetilde{\cal H}= \left( \begin{array}{cc} \dfrac{\hbar^2}{2 m^{\star} R^2} (n- \frac{1}{2})^2 & - \dfrac{ \hbar \alpha}{R} n \\ \\ - \dfrac{\hbar \alpha}{R} n & \dfrac{\hbar^2}{2 m^{\star} R^2} (n+ \frac{1}{2})^2 \end{array} \right).
\end{equation}
Apart from a trivial rigid energy shift, the eigenenergies are simply obtained as 
$$E_{I, II}(n) = \dfrac{\hbar^2}{2 m^{\star} R^2} \left[n^2 \mp n \sqrt{1 + Q_R^2} \right],$$
where the index $I,II$ refers to the two time-reversed channels guaranteed by Kramers' theorem. The corresponding spinorial eigenstates can be found to be 
\begin{equation*}
\Psi^{I}_{n}(\phi)=\mathrm{e}^{i n \phi} \left( \begin{array}{c} \cos{\frac{\gamma}{2}}  \, \mathrm{e}^{-i \phi /2} \\ \\\sin{\frac{\gamma}{2}} \, \mathrm{e}^{i \phi /2}  \end{array} \right), 
\end{equation*}
\begin{equation*}
\Psi^{II}_{n}(\phi)=\mathrm{e}^{i n \phi} \left( \begin{array}{c} \sin{\frac{\gamma}{2}}  \, \mathrm{e}^{-i \phi /2} \\\\ -\cos{\frac{\gamma}{2}} \, \mathrm{e}^{i \phi /2} \end{array} \right).
\end{equation*}
Here the tilt angle $\gamma$ is related to the dimensionless Rashba strength $Q_R$ introduced above by $\tan{\gamma}=Q_R$. 
\begin{figure}[t]
\begin{center}
\includegraphics[width=.75\columnwidth]{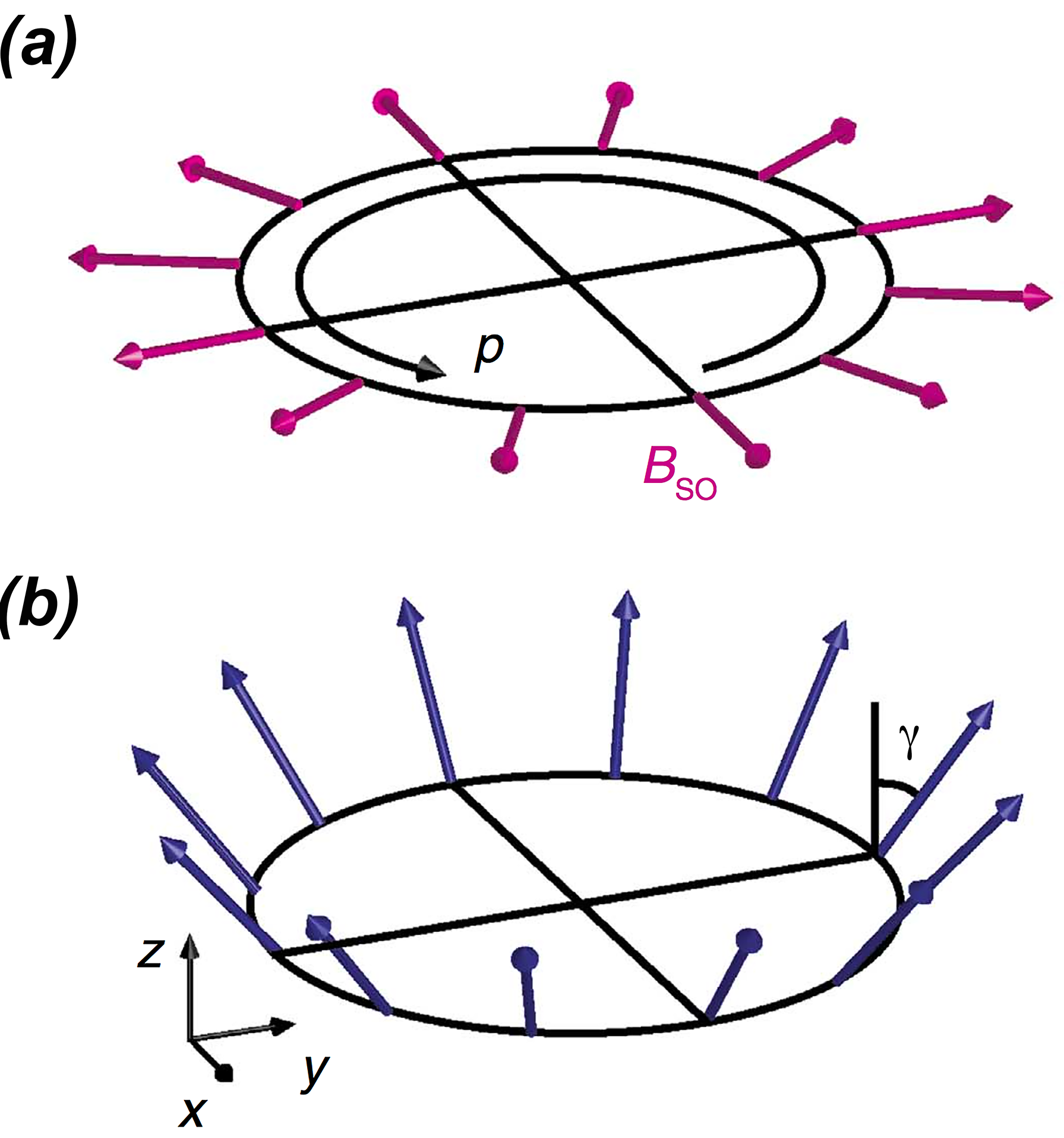}
\end{center}
\caption{(a) Quantum ring with Rashba spin-orbit coupling. The spin-orbit coupling induces an effective in-plane magnetic field $B_{SO}$, which is perpendicular to the electron momentum $p$. (b) In non-adiabatic transport, the electron spin do not align to $B_{SO}$ but acquire an additional out-of-plane component (Adapted from Ref.~\cite{nag13}).}
\label{fig:fig2}
\end{figure}
In the limit of strong Rashba spin-orbit interaction, {\it i.e.} $Q_{R} \rightarrow \infty$, the tilt angle $\gamma \rightarrow \pi/2$ in which case the eigenstates of the Hamiltonian correspond to the spin eigenstates of $\sigma_N$. This limit therefore corresponds to the ``adiabatic" limit in which the spin carriers of the quantum ring orient along the effective momentum dependent Rashba magnetic field in the in-plane normal direction [c.f. Fig.~\ref{fig:fig2}(a)]. For finite values of the dimensionless Rashba strength $Q_R$ instead, the spin carriers acquire a finite out-of-plane component, which is a unique signature of the non-adiabatic spin transport along the ring [c.f. Fig.~\ref{fig:fig2}(b)]. 
Such a non-adiabaticity in the spin motion is immediately reflected in the ballistic transport. Considering as before, incoming spins that propagate coherently along the four available channels of the quantum ring, {\it i.e.} $\ket{n_1, I}$; $\ket{n_2, II}$; $\ket{-n_1, II}$; $\ket{-n_2,I}$, we have that the mixed spin state leaving the ring at $\phi=\pi$ can be written as 
$$\ket{\sigma_{out}} = \sum_{s= I,II} \sum_{i=1,2} \braket{\Psi_{n_i}^s(\phi=0) | \sigma} \times \ket{\Psi_{n_i}^s(\phi=\pm \pi)}. $$
where $\pi$ ($-\pi$) refers to the modes propagating along the upper branch and the lower branch of the quantum ring respectively. 
By summing the spin-resolved quantum transmission probabilities, we obtain that the total conductance takes the following form
\begin{equation}
G= \dfrac{e^2}{h} \left[1 - \cos{(n_1 - n_2) \pi}\right]
\label{eq:conductance2}
\end{equation}
From the eigenenergies written above, we have that $n_1-n_2 = \sqrt{1+ Q_{R}^2}$ and thus the total conductance can be written as 
\begin{equation}
G= \dfrac{e^2}{h} \left[1 + \cos{ \left(\pi \sqrt{1+Q_R^2} - \pi \right)}\right]
\label{eq:conductanceringnormal}
\end{equation}
There are two features that differentiate the conductance oscillations in Eq.~\ref{eq:conductanceringnormal} as compared to the oscillations predicted for a quantum ring with a spin-orbit coupling originating from a radial electric field. First, contrary to the uniform oscillations found in Eq.~\ref{eq:conductance1}, Eq.~\ref{eq:conductanceringnormal} implies the occurrence of quasiperiodic oscillations for small Rashba strength $Q_R<1$. 
Second, in the large Rashba regime $Q_{R} \gg 1$, one observes a relative $\pi$ phase shift between the two conductance modulations . 

As a spoiler for the next section, we here anticipate that this $\pi$ phase shift is the principal consequence of the $\pi$ Berry phase ~\cite{berry} acquired by the spins while precessing around the effective momentum dependent radial Rashba magnetic field due to the out-of-plane electric field.  
\begin{figure}[t]
\begin{center}
\includegraphics[width=.75\columnwidth]{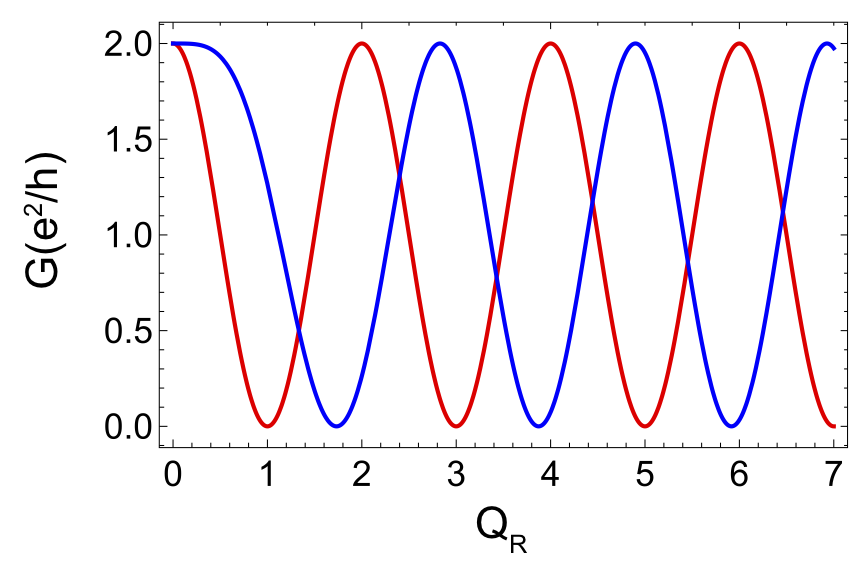}
\end{center}
\caption{Conductance modulation profiles of one-dimensional quantum rings as a function of the dimensionless Rashba strength $Q_{R}$. The blue line corresponds to a Rashba spin-orbit originating from an electric field orthogonal to the ring plane, whereas the red line is for a Rahsba spin-orbit due to a radial electric field. The latter also corresponds to the incomplete result of Ref.~\cite{nit99} .The conductance modulation profiles agree with a related model for one-dimensional rings based on a transfer matrix approach~\cite{mol04}. }
\label{fig:fig3}
\end{figure}
Furthermore, the specific influence of quantum geometric phases in the conductance can be also seen by rewriting Eq.~\ref{eq:conductanceringnormal} as follows
$$G= \dfrac{e^2}{h} \left\{1 + \cos{\left[\pi Q_R \sin{\gamma} -\pi\left(1-\cos{\gamma}\right) \right]} \right\}.$$
The phase in the equation above has then two important contributions: One is the dynamical phase $\pi Q_R \sin{\gamma}$ who also manifests itself for a radial electric field. The other is the Aharonov-Anandan~\cite{aha87} phase $\pi\left(1-\cos{\gamma}\right)$ for non-adiabatic cyclic motion. It corresponds to the solid angle accumulated by the change of spinor orientation during transport, and reduces to the $\pi$ spin Berry phase in the purely adiabatic limit $\gamma \rightarrow \pi/2$. 
This formulation of the conductance in terms of geometric and dynamical phases will be analyzed in detail in the next section.

\section{Conductance modulations as a probe of the Aharonov-Anandan geometric phase}
In this section, we derive the relation between the conductance modulation and the Aharonov-Anandan geometric phase~\cite{aha87}  for a quantum ring with generic shape. This will also allow us to show that real-space geometric deformations directly influence the geometric quantum phase and hence the spin transport properties.  

We start out from the  one-dimensional Hamiltonian in the presence of Rashba spin-orbit interaction (due to a perpendicular electric field) derived in Sec.~\ref{sec:hamiltonian}:
\begin{equation}
{\cal H}= -\dfrac{\hbar^2}{2 m^{\star}} \partial_s^2 + \dfrac{i \hbar \alpha}{2} \left[\sigma_N(s) \partial_s + \partial_s \sigma_N(s)\right],
\label{eq:hamiltoniankp}
\end{equation}
where, for simplicity, we have disregarded the quantum geometric potential since it can be assumed to be a small perturbation as compared to the Rashba spin-orbit interaction. 
Let us discuss the spin textures that are generally realised in a quantum ring with Rashba spin-orbit interaction. To show this, we rewrite the Hamiltonian written above as 
\begin{eqnarray*}
{\cal H}=H^2_l-\frac{\alpha^2 m^{\star}}{2} \sigma_0
\end{eqnarray*}
\noindent with $\sigma_0$ being the identity matrix and $H_l$ reading:
\begin{eqnarray*}
H_l=\left( i \dfrac{\hbar}{\sqrt{2 m^{\star}}} \partial_s+ \frac{\alpha \, \sqrt{m^{\star}}}{\sqrt{2}} \sigma_N(s) \right).
\end{eqnarray*}
Clearly, $H_l$ and ${\cal H}$ have common eigenstates with an eigenvalue relation given by $E=E^2_l- \alpha^2 m^{\star} / 2$.
Let us now introduce the spin orientation of a given spin eigenmode $|\Psi _{E}\rangle$ as the corresponding expectation value
of the spin operators in the local Frenet-Serret reference frame (see Section.~\ref{sec:hamiltonian}), i.e.    
$\langle \boldsymbol{\sigma} \rangle=\{\langle {\sigma}_T \rangle,\langle {\sigma}_N \rangle,\langle {\sigma}_z \rangle \}$.
It is possible to determine the equation for the spatial derivative of the local spin components using that 
the Schr\"odinger equation  $H_l |\Psi _{E}\rangle =E_l |\Psi _{E}\rangle$ can be rewritten as 
\begin{eqnarray}
i \partial_s |\Psi _{E}\rangle &&= \hat{G}(s) |\Psi _{E}\rangle \label{evolve} \\
 \langle \Psi _{E}| i \partial_s&&= - \langle \Psi _{E}| \hat{G}(s) \nonumber
\end{eqnarray}
where we introduced the operator $\hat{G}(s)$
$$\hat{G}(s)=-  \dfrac{\sigma_N(s)}{2 \, l_{\alpha}} - \sigma_0 \sqrt{\dfrac{2 m^{\star} E}{\hbar^2} + \dfrac{m^{\star \,^2} \alpha^2}{\hbar^2}},$$
and $l_{\alpha}$ is the characteristic spin-orbit interaction length defined by $1/l_{\alpha}= 2 m^{\star} \alpha / \hbar$.  
Eqs.~\ref{evolve} yield the general   
expression for the spatial derivative of the expectation value of the spin components
\begin{eqnarray}
\partial_s \langle \boldsymbol{\sigma} \rangle = i\langle [G, \boldsymbol{\sigma}]\rangle + \langle \partial_s \boldsymbol{\sigma} \rangle \,
\label{eq:deriv}
\end{eqnarray}
\noindent with $[A,B]$ indicating the commutator of $A$ and $B$. Using the commutation relations for the local Pauli matrices we have
\begin{equation}
\left\{ \begin{array}{ccc} \left[\hat{G}(s), \sigma_T(s) \right]&=& i \dfrac{\sigma{z}}{l_{\alpha}} \\ \\
 \left[\hat{G}(s), \sigma_N(s) \right]&=& 0 \\ \\
 \left[\hat{G}(s), \sigma_z \right]&=& - i \dfrac{\sigma_T}{l_{\alpha}} 
\end{array}
\right.
\label{eq:commutators}
\end{equation}
\begin{figure}[t]
\begin{center}
\includegraphics[width=.75\columnwidth]{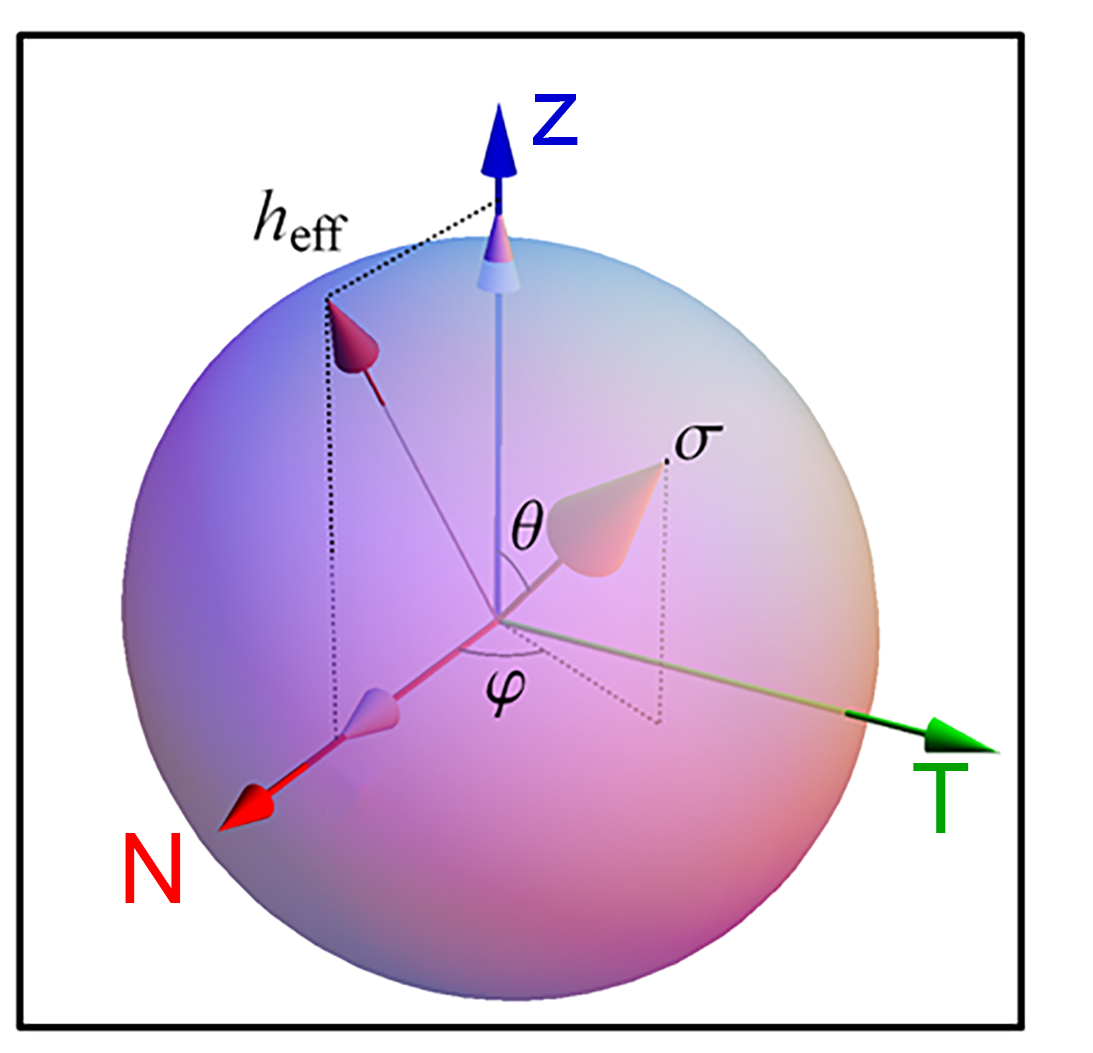}
\end{center}
\caption{The Frenet-Serret-Bloch sphere in the moving frame of the charge carriers in a generic quantum ring with the vectors associated to the electron spin orientation and the effective local field ${\boldsymbol h}_{eff}$. Adapted from Ref.~\cite{yin16}.}
\label{fig:fig4}
\end{figure}
To proceed further, we use that the spatial derivative of the local Pauli matrices obey the Frenet-Serret equations, $\partial_s {\sigma}_N(s)= - \kappa(s) {\sigma}_T(s)$ and $\partial_s {\sigma}_T(s)= \kappa(s) {\sigma}_N(s)$, with $\kappa(s)$ the local curvature. When combining these relations with Eqs.~\ref{eq:commutators}, we therefore find the following equations for the spatial derivative of the spin expectation values: 
\begin{equation}
\left\{\begin{array}{ccc}
\partial_s \langle \sigma_N \rangle &=& -\kappa(s) \langle \sigma_T \rangle \\
\partial_s \langle \sigma_T \rangle &=& - \dfrac{\langle \sigma_z \rangle}{l_{\alpha}} + \kappa(s) \langle \sigma_N \rangle \\
\partial_s \langle \sigma_z \rangle &=& \dfrac{\langle \sigma_T \rangle}{l_{\alpha}}
\end{array}
\right.
\label{eq:gyro}
\end{equation}
The equations above represent a fundamental relation that links the geometric curvature of the quantum ring, the Rashba SO coupling, and the electron 
spin orientation in the local Frenet-Serret frame. It can be also written in the compact form
\begin{equation}
\partial_s \langle \boldsymbol{\sigma} \rangle = - \boldsymbol{h}_{\text{eff}} \times \langle \boldsymbol{\sigma} \rangle , \label{eq:gyroscope}
\end{equation}
where we introduced the local field 
$\boldsymbol{h}_{\text{eff}}=\{0, l_{\alpha}^{-1}, \kappa(s) \}$ which lies in the normal-binormal plane, and depends on the local curvature and effective spin-orbit length introduced above. With this, it also follows that the spin direction lives in a Frenet-Serret-Bloch sphere~\cite{yin16} [see Fig.~\ref{fig:fig4}].
Eq.~\ref{eq:gyroscope} generally implies that due to a non zero curvature, the electron spin acquires a finite out-of-plane binormal $\hat{z}$ component. 
In particular, for a circular quantum ring where the curvature is constant $\kappa(s)=-1/R$ we find, in agreement with the results presented in Section~\ref{sec:circularring}, a local spin orientation given by $\tan{\theta}=2 m^{\star} \alpha R / \hbar = Q_R$ [c.f. Fig.~\ref{fig:fig4}]. 
More importantly, 
a non trivial component along the tangential direction appears provided the curvature is not constant. Although the derivative $\partial_s$ of the spin vector locally vanishes 
if the spin is aligned to the effective spin-orbit field, variations of the local curvature yields a non-vanishing torque which results into a component of the spin vector parallel to
the electron propagation direction. 
Such a torque effect due to the geometric shape of the quantum ring is manifested by considering the 
example of a quantum ring of total  length $L$ with an elliptical shape and a ratio $a/b$ between the minor ($a$) and 
the major ($b$) axes of the ellipse. 
This is a paradigmatic case of a quantum ring with positive but non-uniform curvature that can be suitably enhanced (suppressed) at the 
positions nearby the poles of the major (minor) axes.
There are two distinct spin texture regimes in this Rashba quantum ring. 
For very strong spin-orbit interactions or quasi-constant curvature, {\it i.e.} $a/b \simeq 1$, the electron spin is pinned nearby the quasi-static effective field $\boldsymbol{h}_{\text{eff}}$ in the Frenet-Serret-Bloch sphere. In the regime of weaker spin-orbit interaction or sizable non-uniform curvature profile, instead, the electron spin is not able to follow the 
periodic motion of the effective spin-orbit field. As a result, a finite spin component along the tangential direction appears, and in the local frame the electron spin starts to wind both around the normal and the binormal directions. These features of the spin textures are shown in Fig.~\ref{fig:fig5} where we report the spin textures of an elliptical quantum ring obtained by solving a tight-binding model Hamiltonian derived by discretizing Eq.~\ref{eq:hamiltoniankp} on an atomic chain~\cite{yin16}. For very weak spin-orbit coupling strength [c.f. Fig.~\ref{fig:fig5}(a)], the spin textures are almost aligned along the binormal direction $\hat{z}$. In an intermediate regime of Rashba spin-orbit strength instead, the torque exerted on the spin yields complex three-dimensional spin textures [c.f. Fig.~\ref{fig:fig5}(b)-(f)]. In the very large spin-orbit interaction regime instead, the spin completely aligns along the normal direction signaling an almost adiabatic spin motion.
\begin{figure}[t]
\begin{center}
\includegraphics[width=\columnwidth]{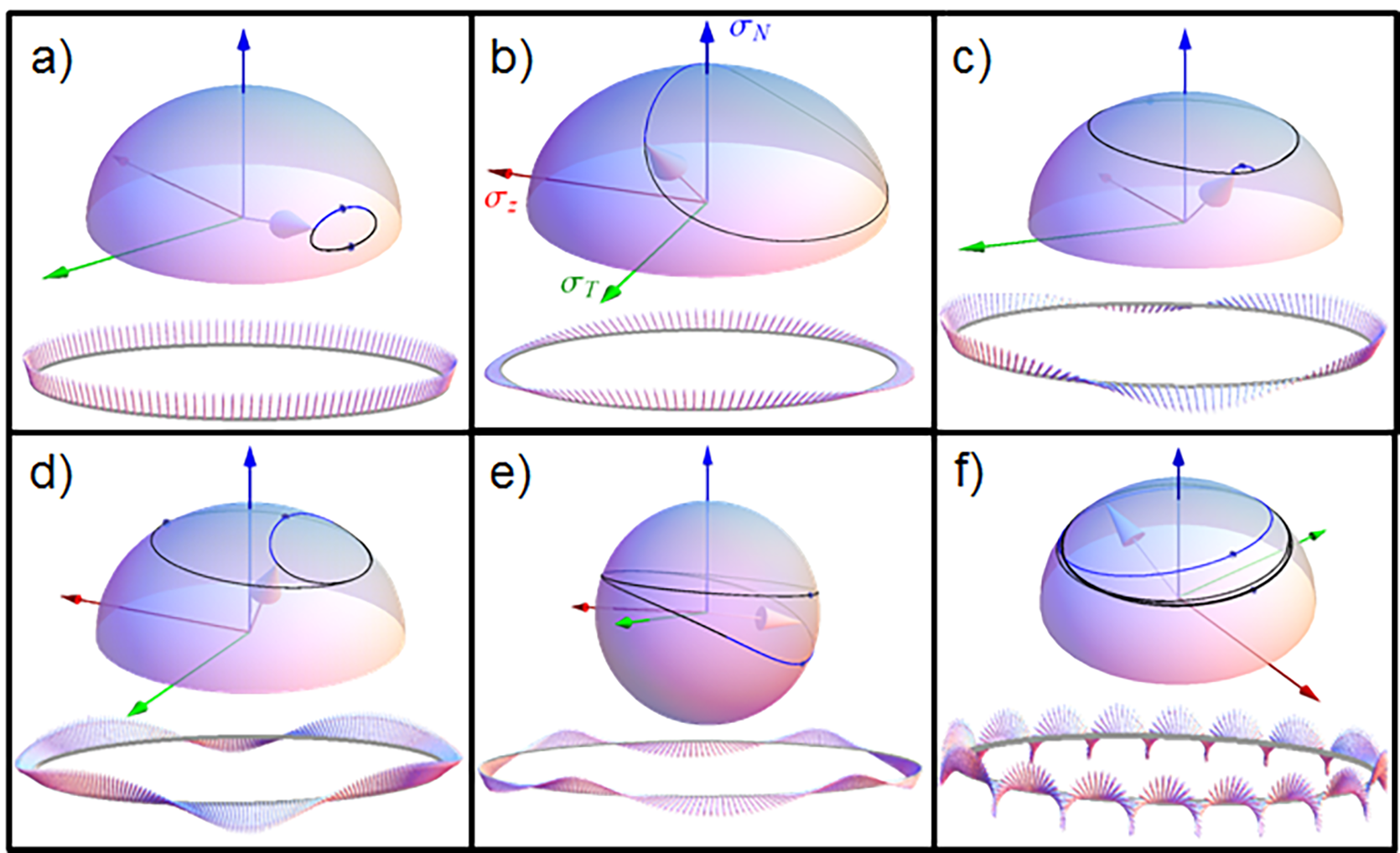}
\end{center}
\caption{Evolution of the electronic trajectories on the Bloch-Frenet-Serret sphere and spin textures in the lab frame for a quantum ring with elliptical shape with ratio between the minor and major axes $a/b=0.4$ and different values of the spin-orbit coupling strength $\alpha$. Panels (a)-(f) correspond to $m^{\star} \alpha L / \hbar = 1,4,8,10,12,50$. From Ref.~\cite{yin16}.}
\label{fig:fig5}
\end{figure}

\begin{figure}[t]
\begin{center}
\includegraphics[width=.85\columnwidth]{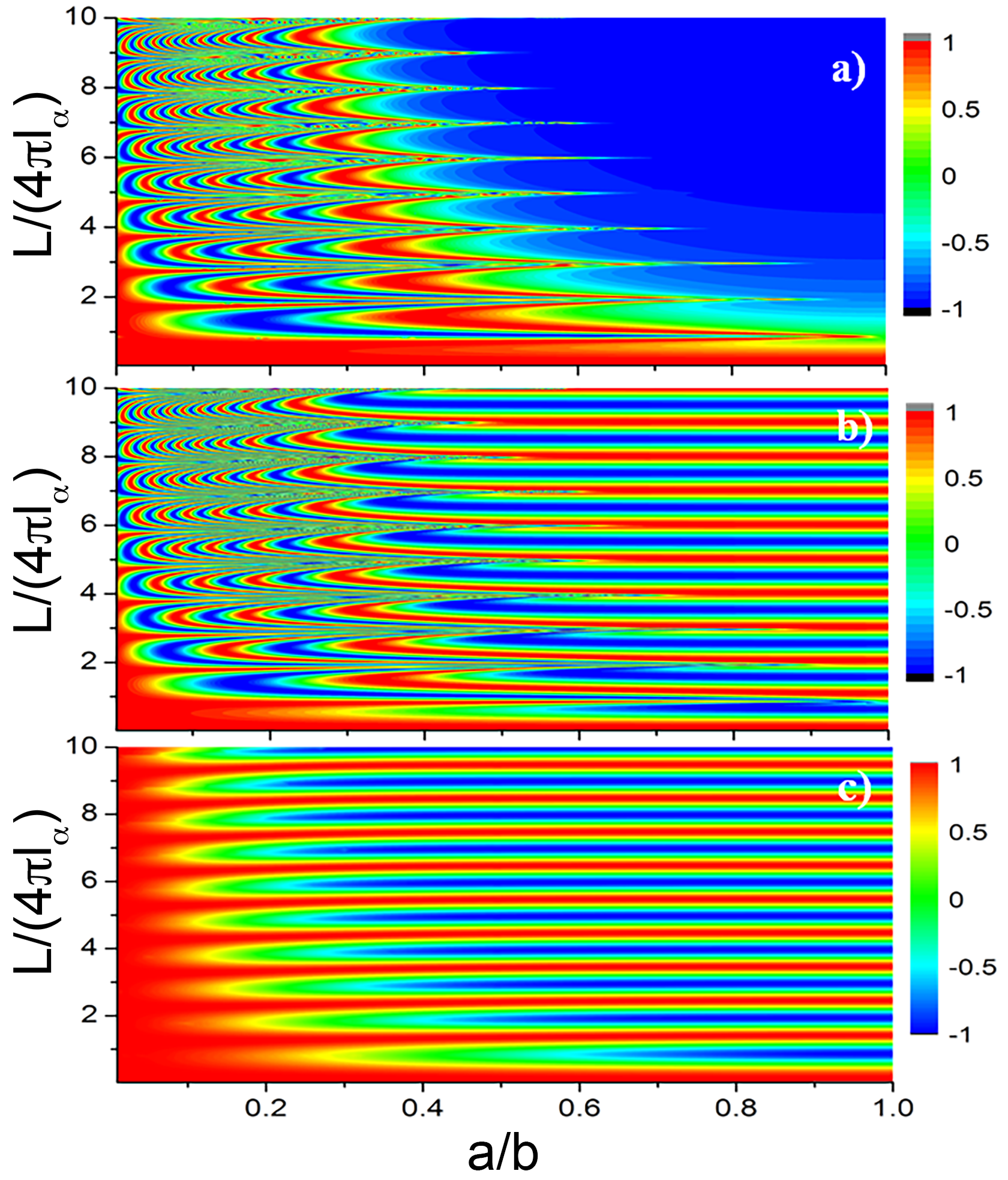}
\end{center}
\caption{Contour map of the cosine of the geometric phase (a), the spin component of the dynamical phase (b), and the total phase (c) contributing to the conductance for a quantum ring of total length $L$ with elliptic shape as a function of the ellipse ration $a/b$ and the dimensionless spin-orbit coupling strength $L/(4 \pi l_{\alpha})$. From Ref.~\cite{yin16}. }
\label{fig:fig6}
\end{figure}

These variety of complex three-dimensional spin textures are not only interesting {\it per se}:  they indeed strongly impact the spin transport properties.  
To show this, we will now find a link between the spin textures and the quantum phases for a cyclic evolution in a generic quantum ring.
We therefore start by noticing that the real space evolution of the spin eigenmode is regulated by Eq.~\ref{evolve}. Closely following Aharonov and Anandan~\cite{aha87}, we use that for any one-dimensional quantum ring the spinorial wavefunction $|\Psi (s)\rangle$ must satisfy the  condition
$$|\Psi ( L)\rangle \equiv \mathrm{e}^{i \chi} |\Psi (0)\rangle.$$
We then define a new wavefunction $|\tilde{\Psi} (s)\rangle =e^{-i \beta(s)} |\Psi (s)\rangle$ in such a way that $\beta(L)-\beta(0)=\chi$. It immediately follows that  
$|\tilde{\Psi} (L)\rangle=|\tilde{\Psi} (0)\rangle$ and from Eq. \ref{evolve} that
\begin{eqnarray*}
-\partial_s \beta(s)= \langle \Psi| \hat{G}(s) |\Psi\rangle- \langle \tilde{\Psi}| i\partial_s |\tilde{\Psi}\rangle\,.
\end{eqnarray*}
Therefore, we can express the total phase $\chi$ accumulated by the charge carriers once they complete the spatial loop as the sum of a geometric Aharonov-Anandan (AA) phase and a dynamical phase as follows
\begin{eqnarray}
g_{AA}= \int_{0}^{L}\langle \tilde{\Psi}| i\partial_s |\tilde{\Psi}\rangle ds \label{eq:gAA} \\ 
d=-\int_{0}^{L}  \langle \Psi | \hat{G}(s) |\Psi\rangle ds\,.
\end{eqnarray}
The dynamical phase can be immediately linked to the expectation value of the local spin as 
\begin{equation}
d= \dfrac{m^{\star} \alpha}{\hbar} \int_0^L \langle \sigma_N(s) \rangle d s + {\it const}.
\label{eq:dynamical}
\end{equation}
In order to find the relation between the local spin expectation value and the geometric (AA) phase we first relate the local normal direction, as well as the tangential one, to the Euclidean coordinates via:
\begin{equation*}
\left\{\begin{array}{ccc}
\hat{N}(s)& = & \cos{\phi(s)} \hat{x} + \sin{\phi(s)} \hat{y} \\ \\
\hat{T}(s)&=& -\sin{\phi(s)} \hat{x} + \cos{\phi(s)} \hat{y},
\end{array}
\right.
\end{equation*}
where $\phi(s)$ is a real-valued function, which is related to the local curvature via the Frenet-Serret equations  yielding 
$$\phi(s) = - \int_{0}^{s} \kappa(s^{\prime}) d s^{\prime}. $$
Next, we express the normalized spinorial eigenfunction in the following general form
\begin{equation*}
|\Psi  \rangle=\left( 
\begin{array}{c}
\exp[-i \phi(s)/2]\, \exp[i \theta_{\Uparrow}(s)] A_{\Uparrow}(s) \  \\ \\
\exp[i \phi(s)/2]\, \exp[i \theta_{\Downarrow}(s)] A_{\Downarrow}(s)%
\end{array}%
\right),
\end{equation*}

where $A_{\Uparrow,\Downarrow}(s)$ are real-valued functions. Such a general expression is convenient since we can express the expectation values of the local spin components in the following form
\begin{equation}
\left\{
\begin{array}{ccc} \langle \sigma_T \rangle & = & 2 A_{\Uparrow}(s) \, A_{\Downarrow}(s) \sin{\left[ \theta_{\Downarrow}(s) - \theta_{\Uparrow}(s) \right]} \\ \\
 \langle \sigma_N \rangle & = & 2 A_{\Uparrow}(s) \, A_{\Downarrow}(s) \cos{\left[ \theta_{\Downarrow}(s) - \theta_{\Uparrow}(s) \right]}  \\ \\ 
  \langle \sigma_z \rangle & = &  A_{\Uparrow}(s) ^2 - A_{\Downarrow}(s) ^2 
\end{array}
\right.
\end{equation}
Furthermore, we have that $\int_{0}^L \kappa(s^{\prime}) d s^{\prime} = 2 \pi N_{\kappa}$ with $N_{\kappa}$ integer for a closed curve. The same holds true for the phase difference $ \theta_{\Downarrow}(s) - \theta_{\Uparrow}(s) $, which acquires a phase shift $2 \pi W$ with $W$ the winding number of the normal and tangential local spin expectation values around the out-of-plane binormal direction, {\it i.e.} $W=\frac{1}{2 \pi} \int_{0}^{L} q_{NT}(s)$
where we introduced 
$$q_{NT}(s)=\dfrac{\langle {\sigma}_N \rangle \partial_s \langle {\sigma}_T \rangle -\langle {\sigma}_T \rangle \partial_s \langle {\sigma}_N \rangle }{\langle {\sigma}_T \rangle^2 + 
\langle {\sigma}_N \rangle^2}.$$
With this, it follows that 
\begin{equation*}
|\tilde{\Psi} \rangle= \left( 
\begin{array}{c}
A_{\Uparrow}(s) \  \\ 
\exp[i \phi(s)]\, \exp[i (\theta_{\Downarrow}(s)-\theta_{\Uparrow}(s))] A_{\Downarrow}(s)%
\end{array}%
\right) \,,
\end{equation*}
and the AA phase can be simply expressed as 
\begin{equation}
g_{AA}=\pi \left(N_{\kappa} + W -\dfrac{1}{2 \pi} \int \langle {\sigma}_z \rangle [\kappa(s) + q_{NT}(s)] ds \right) \,.
\label{eq:aaphase}
\end{equation} 
The knowledge of both the geometric AA phase Eq.~\ref{eq:aaphase} and the dynamical phase Eq.~\ref{eq:dynamical} also allows to express in a straightforward manner the conductance of a generic ballistic one-dimensional ring. By using that the transmission along the arms of ring can be described using a spin rotation operator~\cite{ber05,yin16}, one finds the relation between the conductance and the quantum phases to be given by 
\begin{equation}
G=\dfrac{e^2}{h} \left\{1 + \cos{\left(g_{AA} + d\right)}\right\},
\end{equation}
where the dynamical phase has to be computed disregarding the constant factor in Eq.~\ref{eq:dynamical}.

For a circular quantum ring, the dynamical as well as the AA phases can be easily computed by noticing that $\langle \sigma_N \rangle= \sin{\gamma}$, and $\langle \sigma_z \rangle=  \cos{\gamma}$. By also considering that $N_\kappa=-1$, we therefore find the result for the conductance modulation anticipated in Section~\ref{sec:circularring}, that is 
$$G=\dfrac{e^2}{h} \left[1+ \cos{\left[\pi Q_R \sin{\gamma} - \pi \left(1- \cos{\gamma}\right)\right]} \right].$$

Most importantly, Eqs.~\ref{eq:dynamical},~\ref{eq:aaphase} directly yield a connection between the complex three-dimensional spin textures due to shape deformations and the spin transport properties. This is manifested in Fig.~\ref{fig:fig6} where we show the influence of the geometric shape deformation on the spin interference patterns for the case of elliptical quantum rings~\cite{yin16}. One can observe distinct geometrically driven channels of electronic transport with a changeover from constructive to destructive interference as the ratio between the ellipse axis $a/b$ increases. This results therefore yield a tight connection between the conductance and the character of the spin textures in a Rashba quantum ring. 

\section{Topological transitions in spin interferometers}
\begin{figure}[t]
\begin{center}
\includegraphics[width=.75\columnwidth]{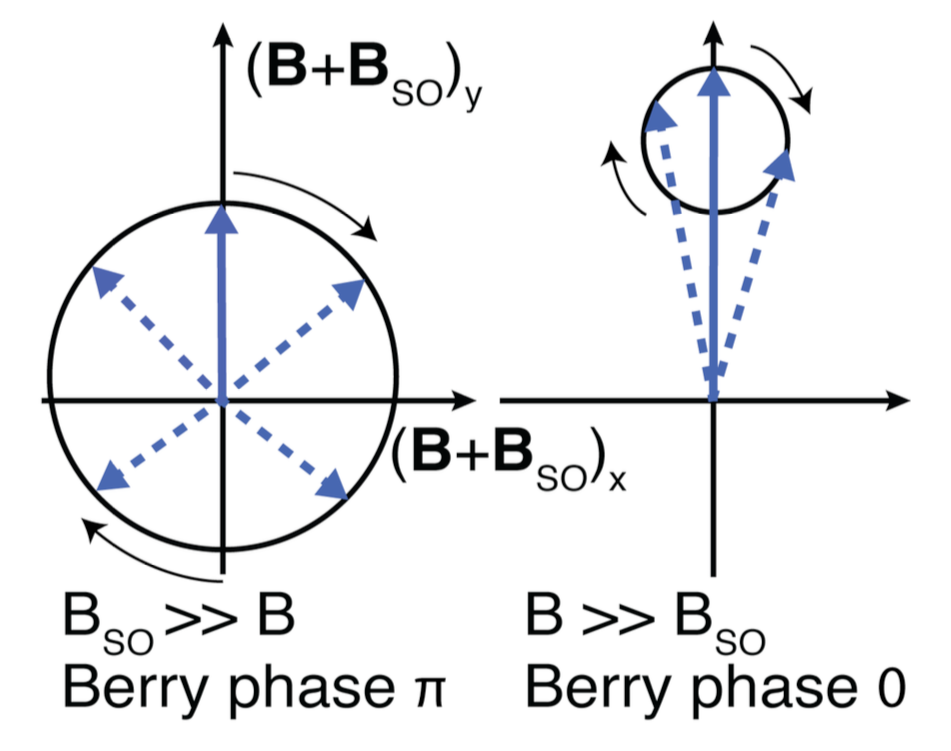}
\end{center}
\caption{The Berry phases in the adiabatic limit for a circular quantum ring with an additional planar magnetic field. For $B_{SO} \gg B$, the accumulated Berry phase correspond to $\pi$. The opposite limit gives instead a $0$ Berry phase. From Ref.~\cite{saa15} (Copyright 2015 American Physical Society) }
\label{fig:fig7}
\end{figure}

In the former section we have shown the connection between the spin textures realized in generic quantum rings and the spin geometric phase, with the latter that can be directly probed by changes in the conductance interference patterns. The spin textures of quantum rings can be also directly controlled using an externally applied magnetic field in the ring plane. The Zeeman coupling
$${\cal H}_Z=g^{\star} \mu_B \sigma_y$$
indeed changes the  solid angle accumulated by the spin eigenmode during transport in a quantum ring and consequently the non-adiabatic AA phase. This can be verified in the small $B$ limit, in which case, by employing standard perturbation theory~\cite{nag13}, the conductance modulations of a circular quantum ring can be written as 
$$G=\dfrac{e^2}{h} \left\{1+ \cos{\left[\pi\left(\sqrt{1+Q_R^2} -1 + \phi(B) \right) \right]}\right\},$$
where $\phi(B) \propto B^2$. This magnetic-field-induced shift in the interference pattern has been experimentally verified in arrays of InGaAs-based quantum rings~\cite{nag13}.
Note that the magnetic field contribution to the conductance modulations only enters in the AA phase. This is because for a quantum ring with symmetrically coupled leads, electronic spins acquire the same Zeeman dynamical phase, and therefore the Zeeman effect only contributes the the geometric part of the quantum phase.

An external magnetic field can, however, also directly modify the topology of the effective magnetic field felt by the carriers during transport, thereby paving the way for the development of topological spin engineering. An early proposal for the topological manipulation of electron spin, which has been put forward by Lyanda-Geller~\cite{lya93}, involved the abrupt switching of Berry phases. Assuming an entirely adiabatic spin transport, it was predicted that a change in the winding number associated with the effective field felt by the charge carriers [c.f. Fig.~\ref{fig:fig7}] would manifest itself as a steplike characteristic in the quantum ring conductance. The intrinsic non-adiabatic nature of the spin transport discussed in the former section, however, requires a more sophisticated approach~\cite{saa15}. 

\begin{figure}[t]
\begin{center}
\includegraphics[width=1.05\columnwidth]{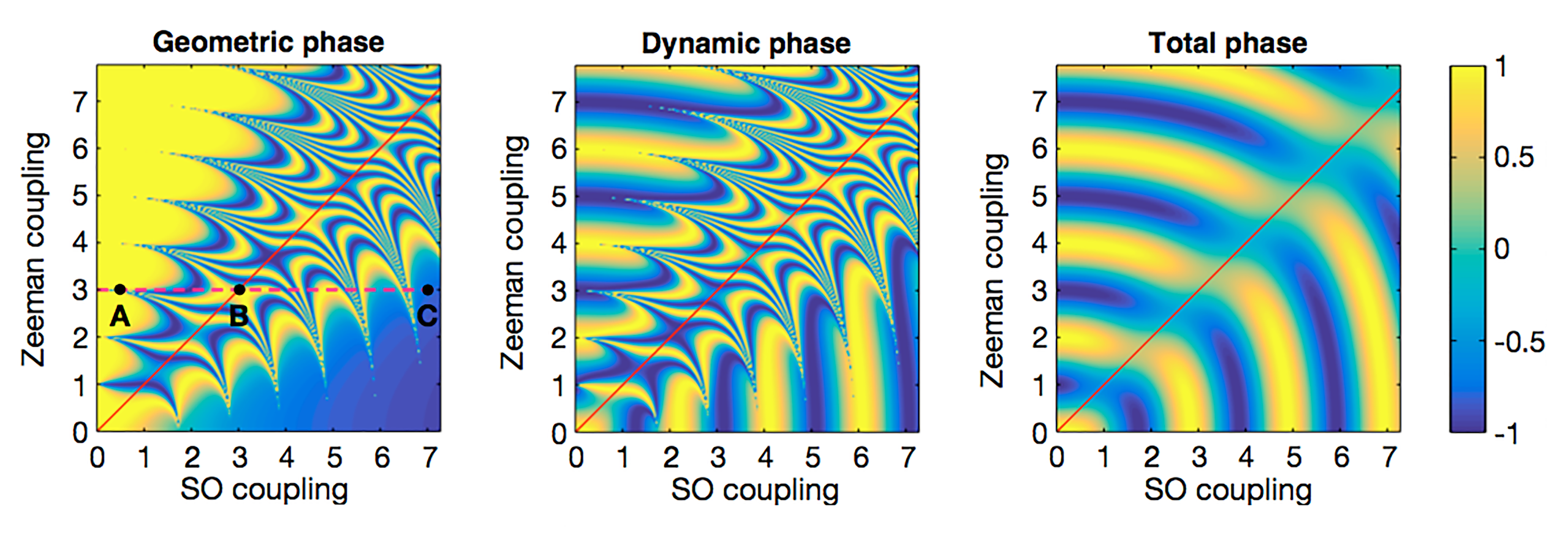}
\end{center}
\caption{Contour map of the cosine of the geometric phase (top panel), the dynamical phase (middle panel), and the total phase (bottom panel) as a function of the spin-orbit coupling and Zeeman splitting in a ballistic single-mode quantum ring tangentially coupled to leads. From Ref.~\cite{saa15} (Copyright 2015 American Physical Society)}
\label{fig:fig8}
\end{figure}

Saarikoski {\it et al.} have thereby analyzed the electronic transport characteristic of a spin interferometer with an externally applied planar magnetic field considering rings tangentially coupled to leads. In this geometric configuration, indeed, the dynamical Zeeman phases can yield both constructive and destructive interference. Henceforth, the conductance will be modulated by both a magnetic field dependent dynamical phase and the magnetic field dependent geometric phase. In Fig.~\ref{fig:fig8} we report the behavior of the two quantum phases in the spin-orbit coupling, magnetic field parameter space. The interference pattern possesses  radial wave fronts, which can be mainly ascribed to Zeeman oscillations. Most importantly, one observes distinct phase dislocations along the critical line where the effective magnetic field textures change topology, {\it i.e.} at $B_{SO}=B$ [c.f. Fig.~\ref{fig:fig7}]. This result is surprising since the topology of the magnetic field textures is reflected in an abrupt change of the conductance modulations even though the spin dynamics is completely non-adiabatic as testified by the complex behaviors of the geometric and dynamic phase for $B_{SO} \simeq B$. 
Whether or not the existence of phase dislocation can be linked to an ``effective" Berry phase with phase slips at the critical line is a matter of future investigations.

\section{Acknowledgements} 
I  acknowledge the financial support from the Future and Emerging Technologies (FET) programme within the Seventh Framework Programme for Research of the European Commission under FET-Open grant number: 618083 (CNTQC), and from a VIDI grant (Project 680-47-543) financed by the Netherlands Organization for Scientific Research (NWO).

\end{document}